\documentclass[twocolumn,showpacs,prb,amsfonts,amsmath,amssymb,floatfix,groupedaddress]{revtex4-2}
\usepackage{color}
\usepackage{hhline}
\usepackage{mathrsfs}
\usepackage{graphicx}
\usepackage{dcolumn}
\usepackage{bm}
\usepackage{multirow}
\usepackage{booktabs}
\usepackage{tabularx}
\usepackage{afterpage}
\usepackage{amsmath}
\usepackage{algorithm}
\usepackage{algorithmic}
\begin{document}

\title{Electronic structure theory of H$_{3}$S: Plane-wave-like valence states, density-of-states peak and its guaranteed proximity to the Fermi level}

\author{Ryosuke Akashi$^{1}$}
\thanks{akashi.ryosuke@qst.go.jp}
\affiliation{$^1$ National Institutes for Quantum Science and Technology (QST), Ookayama, Meguro-ku, Tokyo 152-8550, Japan}

\date{\today}
\begin{abstract}
Superconductivity in sulfur superhydride H$_{3}$S under extreme pressures has been explained theoretically, but it requires a peaked concentration of the electronic density of states (DOS), which has been found in first-principles calculations. The mechanism of this peak formation, though vital for its high transition temperature, has however remained obscure. We address this problem through detailed analysis of the first-principles electronic wave functions. The valence wave functions are shown to be significantly plane-wave-like. From the Fourier-mode analysis of the self-consistent potential and atomic pseudopotentials, we extract the nearly uniform models that accurately reproduce the first-principles band structure with very few parameters. The DOS peak is shown to be the consequence of the hybridization of specific plane waves. Adjacency of Jones' large zone to the plane-wave spherical Fermi surface is posited to be the root cause of the multiple plane-wave hybridization, the DOS peak formation and its proximity to the Fermi level. The present theory resolves the minimal modeling problem of electronic states in H$_{3}$S, as well as establishes a mechanism that may help to boost the transition temperatures in pressure induced superconductors.
\end{abstract}

\maketitle

\section{Introduction}
High-temperature superconductivity by compressing hydrogen-rich compounds~\cite{Ashcroft_PRL1968,Ginzburg_JStatPhys1969,Ashcroft_PRL2004} is a prediction that abuses the formula for superconducting transition temperature ($T_{\rm c}$) from the Bardeen-Cooper-Schrieffer theory~\cite{BCS_PR1957}
\begin{eqnarray}
  T_{\rm c}\propto \omega_{\rm ph}{\rm exp}\left[-\frac{1}{\lambda}\right]
  .
  \label{eq:BCS-Tc}
\end{eqnarray}
The prefactor $\omega_{\rm ph}$ is the representative phonon frequency mediating the electron pairing and $\lambda$ is the effective coupling strength. One could expect high $T_{\rm c}$ by maximizing $\omega_{\rm ph}$ with light elements. Although this scenario seemed rather speculative, it has been made to reality~\cite{Drozdov_Nature2015,Flores_Livas_PhysRep2020,Pickard_AnnRev2021}. A hydrogen-rich phase of the sulfide, H$_{3}$S, is the pioneering example that shows superconducting transition temperature ($T_{\rm c}$) of 200K at pressure more than 100GPa~\cite{Drozdov_Nature2015}. Although the record $T_{\rm c}$ has been soon broken by LaH$_{10}$~\cite{Somayazulu_PRL2019,Drozdov_Nature2019}, YH$_{x}$~\cite{Troyan_AdMa2021,Kong_NatComm2021}, CaH$_{x}$~\cite{LiangMa_PRL2022,LiangMa_PRL2022Err,ZhiwenLi_NatComm2022}, etc., H$_{3}$S is occupying a position of the representative high-$T_{c}$ hydride: Indeed its synthesis has been reproduced by various groups~\cite{Huang_NatSciRev2019,Nakao_JPSJ2019,Osmond_PRB2022}, and various measurements have been applied to this system for characterizing the signature superconductor properties such as the magnetic field responses~\cite{Drozdov_Nature2015,Troyan_Science2016,Mozaffari_NatComm2019,Minkov_NatPhys2023,Minkov_NatComm2022,Minkov_arxiv2024}, isotope effect~\cite{Drozdov_Nature2015,Minkov_AngewChem2020} and superconducting gap~\cite{Capitani_NatPhys2017,Du_Nature2025}.

H$_{3}$S is also attractive from theoretical viewpoints. To date, it is the sole example among the over-200~K family where H is coupled with an atom having larger electronegativity, because of which the celebrated chemical precompression concept~\cite{Ashcroft_PRL2004} may not apply so obviously. Its $T_{\rm c}$ is well reproduced by first-principles calculation based on the Eliashberg theory~\cite{Errea_PRL2015, Akashi_PRB2015,Jose_EPJ2016,Ishikawa_SciRep2016,Akashi_PRL2016,Errea_Nature2016,Sano_PRB2016,Lucrezi_CommPhys2024}. In such calculations a remarkable peak in the density of states (DOS) emerging from the Kohn-Sham (KS) equation has a crucial role~\cite{Papaconstantopoulos_PRB2015,Bianconi_NovSuperMat2015,Bianconi_EPL2015,Jarlborg_SciRep2016,Sano_PRB2016}, as $\lambda$ in the formula Eq.~(\ref{eq:BCS-Tc}) is proportional to the value of DOS at the Fermi level. This peak is easily reproducible, but no simple mechanism as to why this occurs has yet been successfully extracted. Slater-Koster and Wannier-based model constructions~\cite{Bernstein_PRB2015,Ortenzi_PRB2016, Quan_Pickett_PRB2016,Akashi_PRB2020} have revealed that accurate tight-binding modeling requires numerous hopping parameters, implying that the mechanism may be intrinsically complex. The mechanism of the DOS peaking, being the critical factor of the high $T_{\rm c}$, hence remains elusive.

\begin{figure*}[t!]
  \begin{center}
   \includegraphics[scale=0.45]{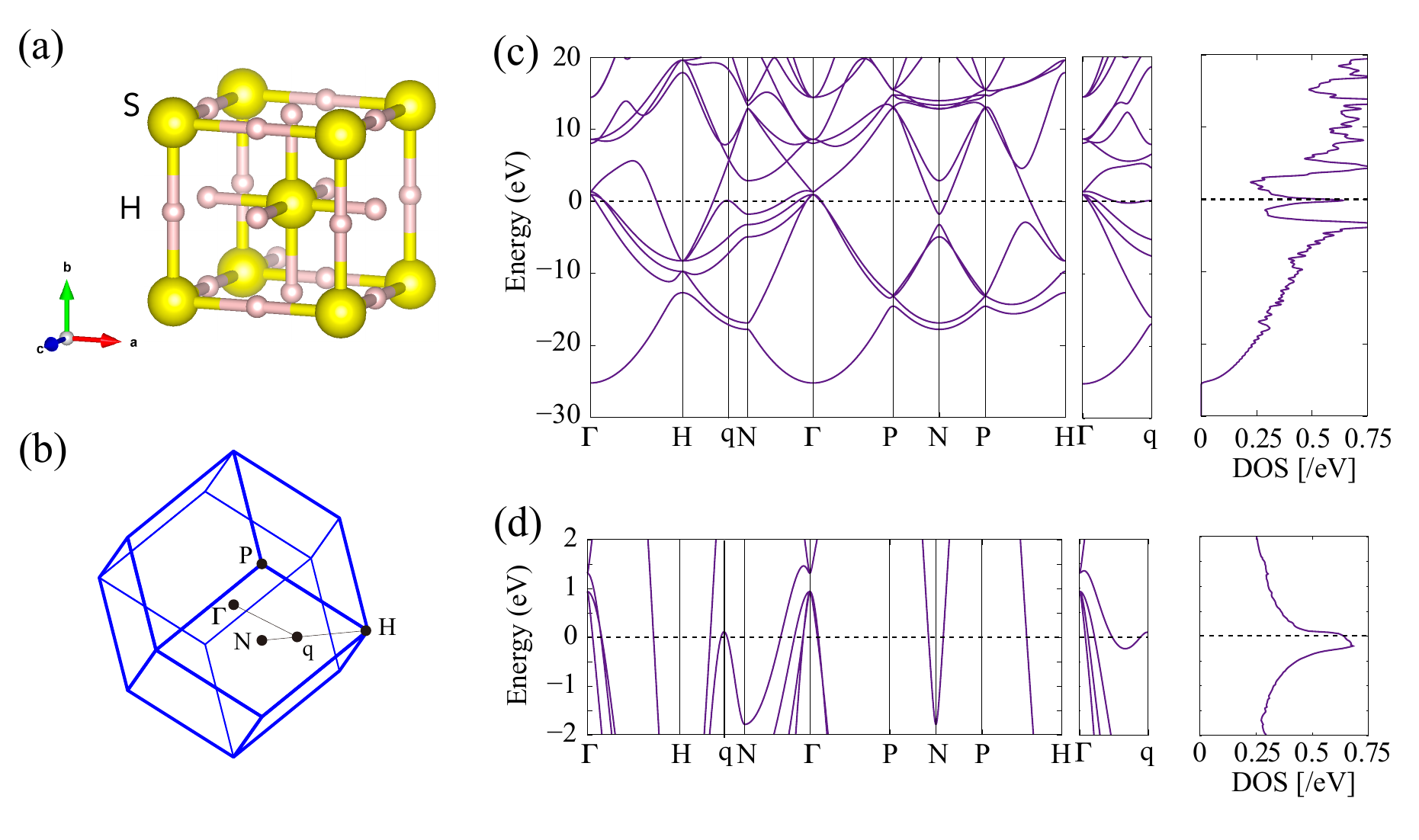}
   \caption{(a) Crystal structure of the high pressure phase of H$_{3}$S, visualized using {\sc VESTA}~\cite{Momma_JApplCrys2011}. (b) The corresponding body centered cubic Brillouin zone. We indicate a point ${\rm q}=(2\pi/a)\times (2/3, 1/3, 0)$ for convenience. (c) Electronic band structures and density of states and (d) their close-up view. The maximum and minimum along $\Gamma$--${\rm q}$ have been found to be saddle points in the other directions~\cite{Akashi_PRB2020,Akashi_Bragg_PRR2024}. The dashed line indicates the Fermi level. } 
   \label{fig:H3S-band-dos}
   \end{center}
 \end{figure*}

In this study, we revisit the minimal modeling problem of the electronic structure in H$_{3}$S from a viewpoint opposite to the tight binding. The author recently pointed out a similarity of its band structure to the uniform electron system~\cite{Akashi_Bragg_PRR2024}. Using the nearly uniform electron model, various essential features in the band structure such as the DOS peak were reproduced with three empirical parameters, suggesting that the band structure formation in H$_{3}$S is understandable by the classic nearly uniform electron theory~\cite{Herman_RMP1958}. We here attempt to derive the emerging nearly uniform model from first principles. We analyze the first-principles electronic wave functions to reveal their plane-wave like charactersistics. The parameters for the nearly uniform model are derived and their optimization toward the first-principles band structure is demonstrated, by which the nearly-uniform modeling as an effective theory is validated. Upon the results, we find that the Jones zone activation mechanism~\cite{Jones_book,Feng_JCP2010} is applicable as a simultaneous solution as to why the DOS peak occurs in the nearly uniform system and why it is pinned near the Fermi level. Successful understanding based on such a general theory dispels the concern that the theoretical DOS peak could be an accidental artifact, and opens a possibility to theoretically design band amonalies near the Fermi level in compressed metallic systems.

\section{Computational detail}
The first-principles electronic band structure of H$_3$S with body-centered cubic structure [Fig.~\ref{fig:H3S-band-dos}(a)] was calculated using with the plane-wave pseudopotential
method as implemented in QUANTUM ESPRESSO~\cite{Giannozzi_JPhysC2017}. The Perdew-Burke-Ernzerhof generalized gradient approximation adapted for solids (PBEsol~\cite{PBEsol_PRL2008}) was adopted for the exchange correlation functional~\cite{Kohn_Sham_PR1965}. The optimized norm-conserving Vanderbilt pseudopotentials~\cite{Hamann_PRB2013} for H and S atoms as provided from {\sc Pseudo Dojo}~\cite{VanSetten_CPC2018} were used. The plane-wave cutoff for the wave function was set to 90 Ry. The charge density was converged self-consistently using the 8$\times$8$\times$8 ${\bf k}$-points. The cubic lattice parameter $a$ was set to 5.6367 bohrs, with which the calculated theoretical pressure amounts to 195 GPa. 

\section{Result}
For later discussions, we first reiterate the established first-principles band structure of H$_{3}$S in Fig.~\ref{fig:H3S-band-dos}. According to the structure search this system stabilizes in the body-centered cubic crystal structure~\cite{Duan_SciRep2014}, the sulfur positions of which were confirmed by x-ray diffraction experiment~\cite{Einaga_NatPhys2016}. The DOS shows the celebrated peak at the Fermi level as shown in panel (c). The peak edges correspond to the energetically close van Hove singularities~\cite{vanHove_PR1953,Quan_Pickett_PRB2016}. This peak, which occupies approximately half of the total DOS at $E_{\rm F}$, is contributed by the Bloch states in tiny linear regions in the ${\bf k}$ space~\cite{Akashi_PRB2020,Akashi_Bragg_PRR2024}. In the previous work~\cite{Akashi_Bragg_PRR2024}, we proposed that this concentration originates from the hybridization of multiple plane waves that are degenerate along a kind of intersections of the Bragg planes. We show the band dispersion along such an intersection in panels (c) and (d), which passes through $\Gamma$ and a low-symmetry point ${\rm q}\equiv(2/3,1/3,0)$. Indeed we see two adjacent saddle points corresponding to the DOS peak edges [Fig.~\ref{fig:H3S-band-dos}(d)]. By introducing a diffraction potential that hybridizes the degenerate plane waves, the nearly uniform model was found to reproduce the overall appearance of the band structure~\cite{Akashi_Bragg_PRR2024}. Below, we examine the first-principles KS wave functions to validate the plane-wave character of the electronic states in this system.

\begin{figure}[t!]
  \begin{center}
   \includegraphics[scale=0.48]{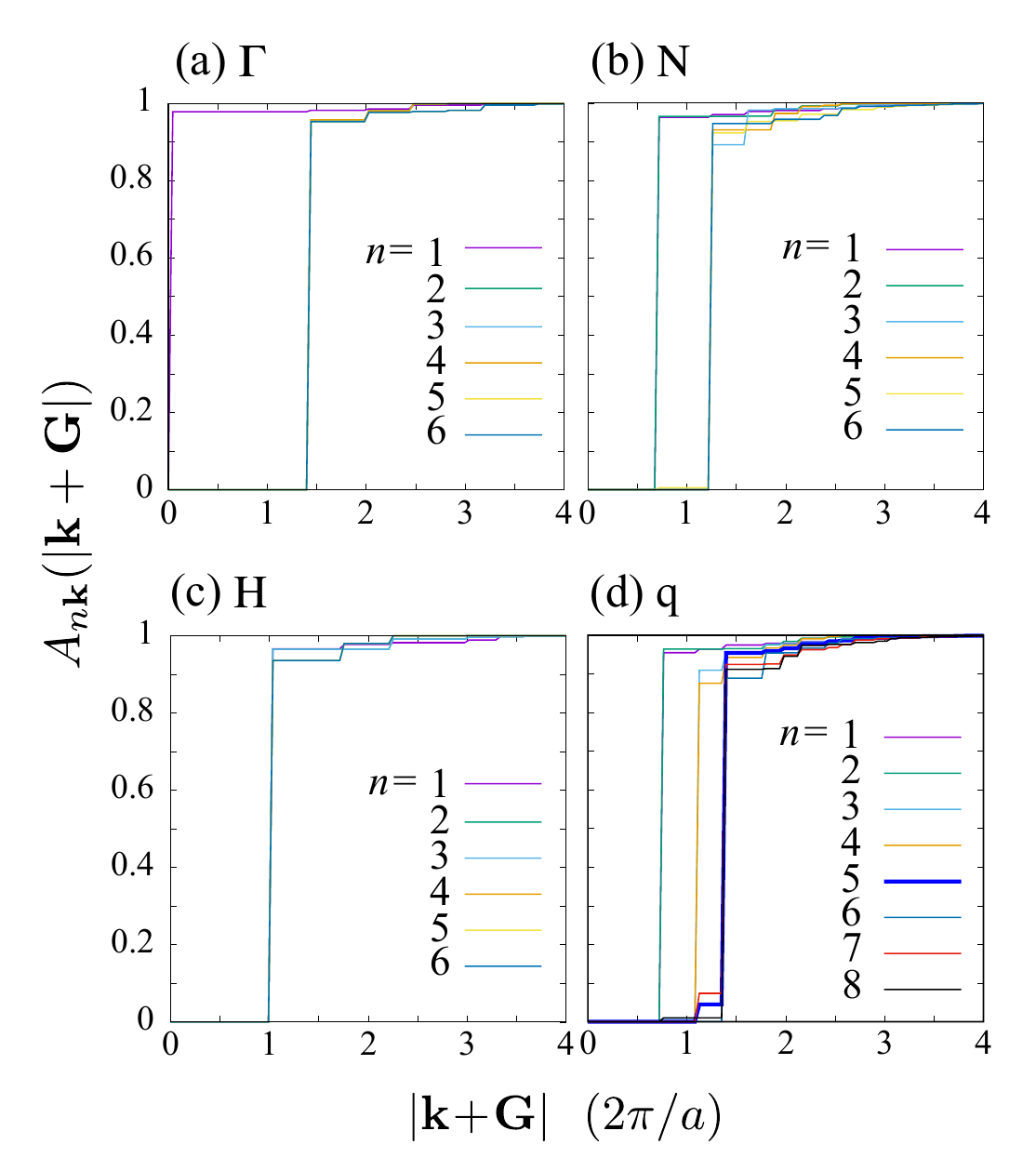}
   \caption{Plane wave components accumulated up to absolute wave number $|{\bf k}+{\bf G}|$ at selected points in the BZ. The band indexes are from lower to higher Kohn-Sham energy eigenvalues. Label ``band 5" at ``q'' corresponds to the band that is responsible for the DOS peak.} 
   \label{fig:H3S-accum-component}
   \end{center}
 \end{figure}

 \begin{figure*}[t!]
  \begin{center}
   \includegraphics[scale=0.50]{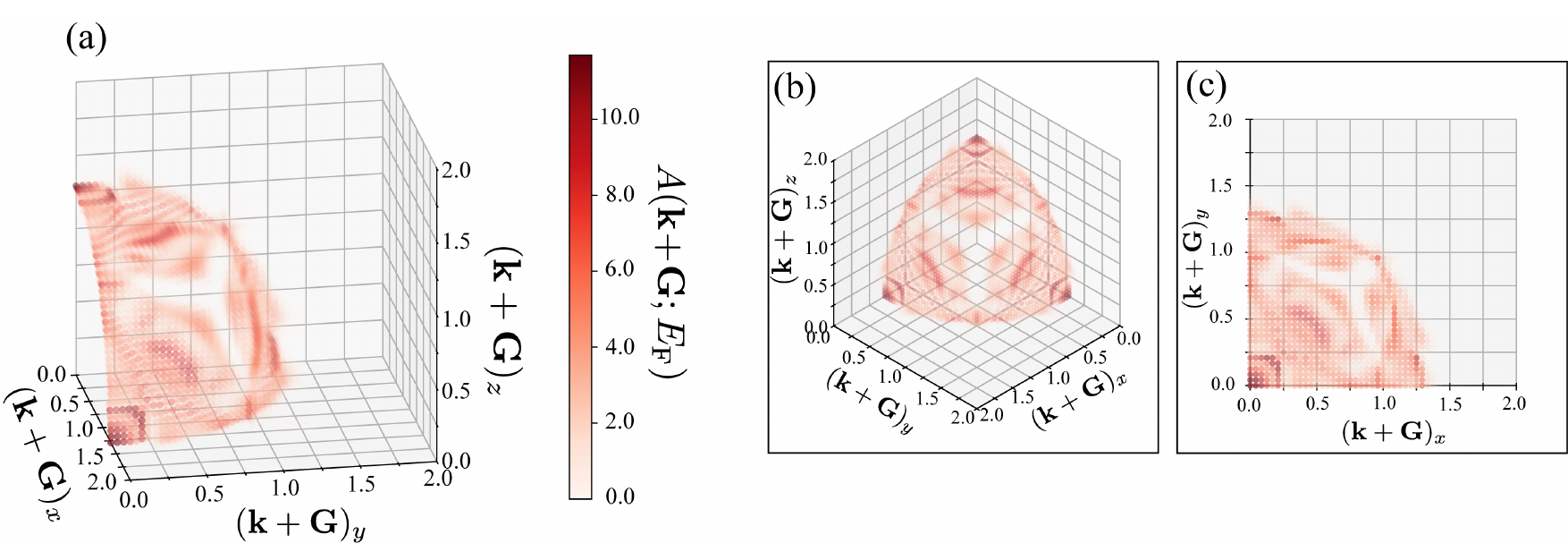}
   \caption{Plane-wave resolved spectral function $A({\bf k}+{\bf G};E_{\rm F})$ plotted in the region $({\bf k}+{\bf G})_{i}\geq 0 \ (i=x,y,z)$. The delta function $\delta$ in Eq.~(\ref{eq:spectral}) was approximated to be Lorentzian with width=0.02 Hartree. We set the opacity of each point to be proportional to the function value. (a) Bird's eye view, (b) views from $(1 1 1)$ axis and (c) from $(0 0 1)$ axis.} 
   \label{fig:H3S-k-spectr}
   \end{center}
 \end{figure*}

\subsection{Kohn-Sham state analysis}

The KS state~\cite{Kohn_Sham_PR1965} obeying the Bloch theorem is labeled by the band index $n$ and crystal wave number ${\bf k}$ and is written as
\begin{eqnarray}
  \psi_{n{\bf k}}({\bf r})
  =
  \sum_{\bf G}c_{n{\bf k}}({\bf k}+{\bf G})e^{i{({\bf k}+{\bf G})\cdot {\bf r}}},
\end{eqnarray}
where ${\bf G}$ is the reciprocal wave number for the corresponding crystal lattice. Using the plane-wave coefficient $c_{n{\bf k}}({\bf k}+{\bf G})$, we calculated
\begin{eqnarray}
  A_{n{\bf k}}(|{\bf k}+{\bf G}|)
  =\sum_{{\bf G}' {\rm s.t.} |{\bf k}+{\bf G}'|<|{\bf k}+{\bf G}|}|c_{n{\bf k}}({\bf k}+{\bf G}')|^2,
\end{eqnarray}
which reveal how the plane waves contribute to form the KS wave function. In Fig.~\ref{fig:H3S-accum-component}, we show $A_{n{\bf k}}$ for several lowest-energy states at selected ${\bf k}$-points. Almost all the lowest energy states showed discontinuous increase of $\gtrsim$ 0.9 at certain definite $|{\bf k}+{\bf G}|$, which indicate that the corresponding states are approximately single plane waves or mixtures of those with degenerate $|{\bf k}+{\bf G}|$. This feature is in stark contrast to typical molecular solids, where local bonds are dominated by localized orbitals which show broader distributions of $|c_{n{\bf k}}({\bf k}+{\bf G})|^2$.

The plane-wave aspect of the electronic states near the Fermi level can also be seen by unfolding the Kohn-Sham states to real momentum by the spectral function
\begin{eqnarray}
  A({\bf k}+{\bf G};E)=\sum_{n}|c_{n{\bf k}}({\bf k}+{\bf G})|^2 \delta(E-\varepsilon_{n{\bf k}}),
  \label{eq:spectral}
\end{eqnarray}
which display the plane-wave modes that contributes to the electronic states at energy $E$. $\varepsilon_{n{\bf k}}$ denotes the KS energy eigenvalue. We plot this quantity at $E=E_{\rm F}$ in Fig.~\ref{fig:H3S-k-spectr}. Overall spectra are concentrated on an approximately spherical surface region. This result signifies that the first-principles KS states around the Fermi level, though apparently complicated in the reduced Brilloin zone expression, are roughly formed by folding the free-electron band $E_{0}({\bf k})=k^2/2$.

 \begin{figure*}[t!]
  \begin{center}
   \includegraphics[scale=0.60]{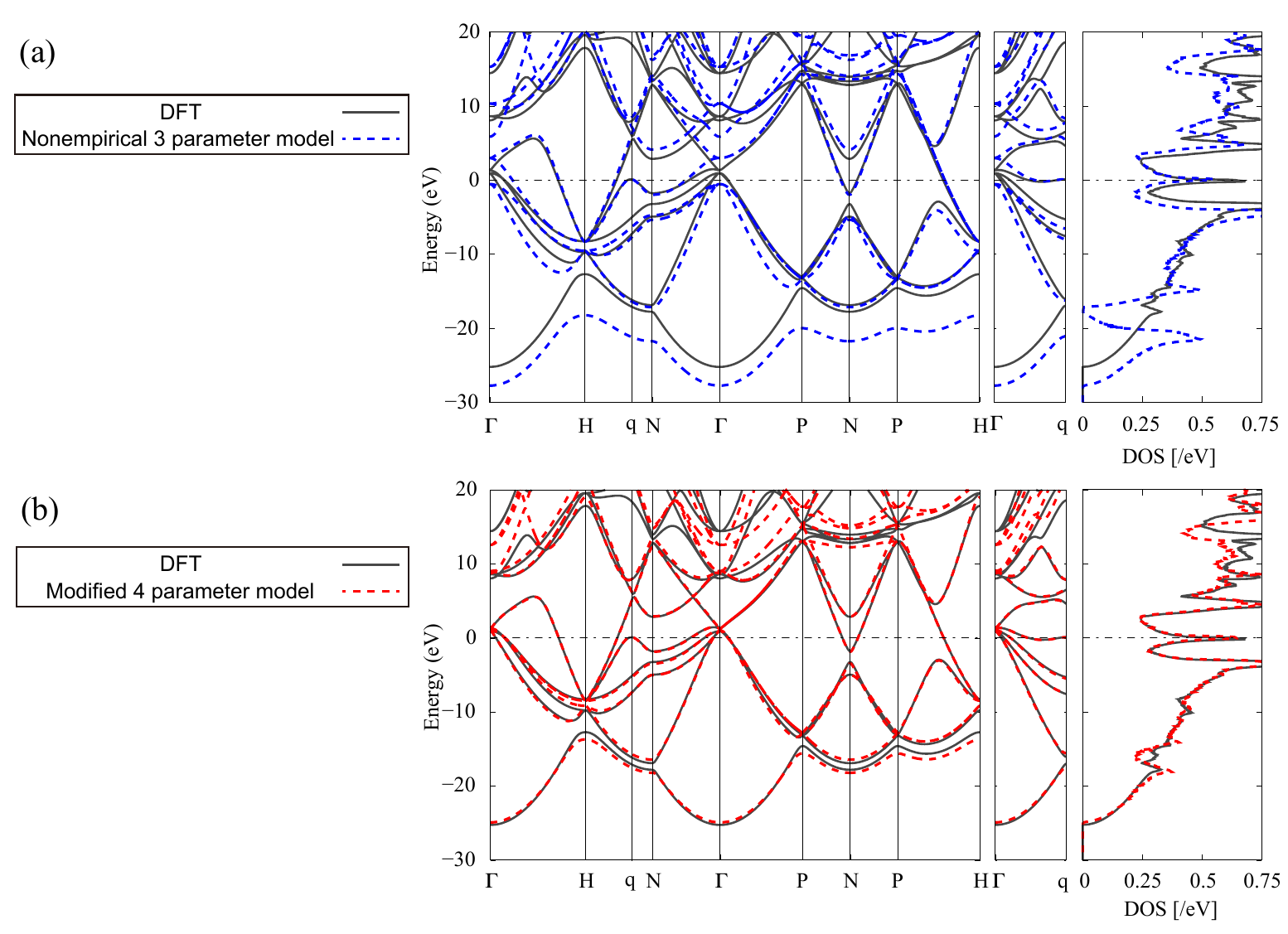}
   \caption{Band structure calculated from first principles compared with effective nearly uniform model results. The Fermi levels of the model results are determined to best fit the band concerning the DOS peak. (a) Comparison with the three-parameter model extracted from the first-principles potentials and (b) with the four-parameter model manually optimized to fit the first-principles band structure.} 
   \label{fig:Bands-compare}
   \end{center}
 \end{figure*}

\subsection{Extraction of the plane-wave model}
We next attempt to formulate a simple nearly uniform model with the plane waves as a basis. The KS band structure in the pseudopotential method is calculated from the one-body KS Hamiltonian including both the local and nonlocal potentials, but the empirical model calculation in Ref.~\cite{Akashi_Bragg_PRR2024} suggests that effective modeling with local potential is possible. We here analyze the first-principles potential components to derive such a model with quantitative accuracy.

The KS Hamiltonian~\cite{Kohn_Sham_PR1965} converged through the self-consistent cycle is as follows:
\begin{eqnarray}
  H=-\frac{\nabla^2}{2}+V_{\rm ion}+V_{\rm Hxc}.
\end{eqnarray}
The Hartree-exchange-correlation part of the potential $V_{\rm Hxc}$ is local, whereas the ionic pseudopotential is composed of local and non-local terms: $V_{\rm ion}({\bf r},{\bf r}')=V_{\rm ion, L}({\bf r})\delta_{{\bf r},{\bf r}'}+V_{\rm ion, NL}({\bf r},{\bf r}')$. Since the total Hamiltonian has the discrete translation symmetry of the crystal, the local potentials are decomposed into the Fourier modes
\begin{eqnarray}
V_{\rm Hxc}({\bf r})=\sum_{{\bf G}}e^{i{\bf G}\cdot {\bf r}}\tilde{V}_{\rm Hxc}({\bf G}),  \\
V_{\rm ion, L}({\bf r})=\sum_{{\bf G}}e^{i{\bf G}\cdot {\bf r}}\tilde{V}_{\rm ion, L}({\bf G}),
\end{eqnarray}
whereas the nonlocal potential is represented by double Fourier decomposition form:
\begin{eqnarray}
  V_{\rm ion, NL}({\bf r},{\bf r}')
  \!=\!\sum_{{\bf k}{\bf G}{\bf G}'}e^{i({\bf k}+{\bf G})\cdot {\bf r}}\tilde{V}_{\rm ion, NL}({\bf k};{\bf G},{\bf G}')e^{-i({\bf k}+{\bf G}')\cdot {\bf r}'}
  .
  \nonumber \\
\end{eqnarray}
The Hamiltonian is block-diagonalized by the crystal wavenumber and its ${\bf k}$-block in the plane-wave representation reads
\begin{eqnarray}
  \setlength{\arraycolsep}{5pt}
  H_{\bf k}=\begin{pmatrix}
      E_{0}({\bf k}-{\bf G}_{0}) & V_{\bf k}({\bf G}_{0},{\bf G}_{1}) & V_{\bf k}({\bf G}_{0},{\bf G}_{2}) & \cdots\\
      V_{\bf k}({\bf G}_{1},{\bf G}_{0}) & E_{0}({\bf k}-{\bf G}_{1}) & V_{\bf k}({\bf G}_{1},{\bf G}_{2}) & \cdots\\
      V_{\bf k}({\bf G}_{2},{\bf G}_{0}) & V_{\bf k}({\bf G}_{2},{\bf G}_{1}) & E_{0}({\bf k}-{\bf G}_{2})  & \cdots\\
      \vdots & \vdots & \vdots&\ddots 
    \end{pmatrix}
    \nonumber \\
    \label{eq:Hamil-k}
\end{eqnarray}
with $E_{0}({\bf k})=k^2/2$ and $\{{\bf G}_{0},{\bf G}_{1},\cdots\}$ are the reciprocal vectors. The non-diagonal matrix element is given by 
\begin{eqnarray}
  &&V_{\bf k}({\bf G},{\bf G}')
  =\tilde{V}_{\rm Hxc}({\bf G}-{\bf G}')+\tilde{V}_{\rm ion,L}({\bf G}-{\bf G}')
  \nonumber \\
  && \hspace{80pt}+\tilde{V}_{\rm ion, NL}({\bf k};{\bf G},{\bf G}')
  .
\end{eqnarray}

In the previous work~\cite{Akashi_Bragg_PRR2024}, we have found that by assuming the local potential form $V_{\bf k}({\bf G},{\bf G}')=V({\bf G}-{\bf G}')$ a modeling by only three parameters $\{V_{1},V_{2},V_{3}\}=\{V({\bf G})\}$ with $|{\bf G}|=\sqrt{2}, 2, \sqrt{6}$ yields a band structure surprizingly similar to the first-principles one. Here, we relate this empirical modeling quantitatively with the first-principles self-consistent potential.

Using the self-consistently converged $\tilde{V}_{\rm Hxc}$, we calculated the Fourier components of the local part $\tilde{V}_{\rm Hxc}({\bf G})+\tilde{V}_{\rm ion,L}({\bf G})$ and non-local part $\tilde{V}_{\rm ion, NL}({\bf k};{\bf G},{\bf G}')$. The latter can be calculated from the pseudopotentials used for the KS calculation via
\begin{eqnarray}
  &&\tilde{V}_{\rm ion, NL}({\bf k};{\bf G},{\bf G}') \nonumber \\
  && = \sum_{a}\sum_{{\bf t}_{a}}e^{-i({\bf G}-{\bf G}')\cdot{\bf t}_{a}}\sum_{\mu\nu}\tilde{\beta}_{a\mu}({\bf k}-{\bf G})D^{a}_{\mu\nu}\tilde{\beta}^{\ast}_{a\nu}({\bf k}-{\bf G}').
  \nonumber \\
\end{eqnarray}
The projector function $\beta_{a\mu}$ and matrix $D_{\mu\nu}^{a}$ are read from the pseudopotential file for atom $a=\{{\rm H}, {\rm S}\}$. ${\bf t}_{a}$ are atomic positions of kind $a$ in the unit cell, and $\mu$ and $\nu$ are indices for the projectors. We found that the contribution from H was no more than 3\% of that from S. 

\begin{table}[t]
  \centering
  \caption{Nonempirical diffraction potential parameters in Hartree, extracted from the first-principles calculation. $V_{1}=V({\bf G})$ with $|{\bf G}|=\sqrt{2}$, $V_{2}=V({\bf G})$ with $|{\bf G}|=2$, and $V_{3}=V({\bf G})$ with $|{\bf G}|=\sqrt{6}$.}
  \label{tab:DiffractPot}
  \setlength{\tabcolsep}{8pt}
  \begin{tabular}{cccc}
    \hline
      & $V_{1}$ & $V_{2}$ & $V_{3}$ \\
    \hline
    $V_{\rm ion, L}+V_{\rm Hxc}$ & $-0.1850$ & $-0.1635$ & $-0.0347$ \\
    $V_{\rm ion, NL}$ & +0.1137 & +0.1071 & +0.1049\\
    \hline
    Total & $-0.0713$ & $-0.0564$ & +0.0702 \\
    \hline
  \end{tabular}
\end{table}

Upon these calculations, we derived the abovementioned three parameters by approximating 
\begin{eqnarray}
V_{\rm ion,NL}({\bf k};{\bf G},{\bf G}')\approx V_{\rm ion,NL}({\bf k}_{\rm rep};{\bf G},{\bf G}')
\end{eqnarray} with the representative ${\bf k}$-point ${\bf k}_{\rm rep}$ set to be the midpoint of ${\bf G}$ and ${\bf G}'$. This selection of ${\bf k}_{\rm rep}$ makes $V_{\rm ion, NL}$ local: $V_{\rm ion,NL}({\bf k}_{\rm rep};{\bf G},{\bf G}')=V_{\rm ion,NL}({\bf G}-{\bf G}')$. We display the total potential components with decompositions into contributions from $V_{\rm ion,L}+V_{\rm Hxc}$ and $V_{\rm ion,NL}$ in Table~\ref{tab:DiffractPot}. Using those extracted parameters, we calculated the band structure of Hamiltonian Eq.~(\ref{eq:Hamil-k}) as shown in Fig.~\ref{fig:Bands-compare}(a). The first-principles band structure is reproduced very accurately. In particular, the dispersion in the $\Gamma$--${\rm q}$ path and the resulting peaked DOS near the Fermi level, which is of central interest here, are properly generated. This result indicates that the three parameter extracted from the first-principles calculation properly captures the major characteristics of the electronic structure of H$_{3}$S. 

The remaining deviation from the first-principles band structure can be made smaller by empirically tuning $V_{1}$--$V_{3}$ and modifying the electronic effective mass slightly:
\begin{eqnarray}
  E_{0}({\bf k})\rightarrow E'_{0}({\bf k})=\frac{k^2}{2m^{\ast}}
  .
\end{eqnarray}
With this manipulation we mean to reflect the nonlocality of the sulfur pseudopotential ignored in the current local approximation. The resulting band structure is shown in Fig.~\ref{fig:Bands-compare} (b), which agrees with the first-principles one almost perfectly. The parameters for the calculated band structures are summarized in Table~\ref{tab:ModelParams}, which we claim best represent the electronic structure of the compressed H$_{3}$S.

The decomposed contributions from $V_{\rm ion,L}+V_{\rm Hxc}$ and $V_{\rm ion,NL}$ are insightful for understanding the uniform-electron-like properties of the first-principles electronic states. The negative contributions of the former to $V_{1}$ and $V_{2}$, which represent the totally attractive self-consistent potential, should prefer atomic-like valence wavefunctions. These are however largely cancelled by $V_{\rm ion,NL}$, which flattens the effetive potential felt by electrons and the resulting KS states are rendered delocalized. The latter potential reflects the orthogonalization to the inner core electrons of S ions. The large positive value of $V_{3}$, which strongly hybridizes the plane waves involved~\cite{Akashi_Bragg_PRR2024}, is dominated by $V_{\rm ion,NL}$. The effect of $V_{\rm ion,NL}$ originates from the non-local scattering of the periodically aligned S ions and cannot be inferred from the local potential distribution. Thus, the core electrons of S ions are largely responsible for the plane-wave like valence states as well as the peaked DOS near the Fermi level.

\begin{table*}[t]
  \centering
  \caption{Summary of the nearly uniform model parameters at $a$=5.6367 bohr, in Hartree.}
  \label{tab:ModelParams}
  \setlength{\tabcolsep}{8pt}
  \begin{tabular}{ccccc}
    \hline
      & $V_{1}$ & $V_{2}$ & $V_{3}$ & $m^{\ast}$ \\
    \hline
    Nonempirical $3$ parameter model & -0.0713 & -0.0564 & +0.0702& --- \\
    Modified $4$ parameter model &  -0.0322 & -0.0401 & +0.0715 & 1.087 \\
    \hline
  \end{tabular}
\end{table*}

\begin{figure}[t!]
  \begin{center}
   \includegraphics[scale=0.45]{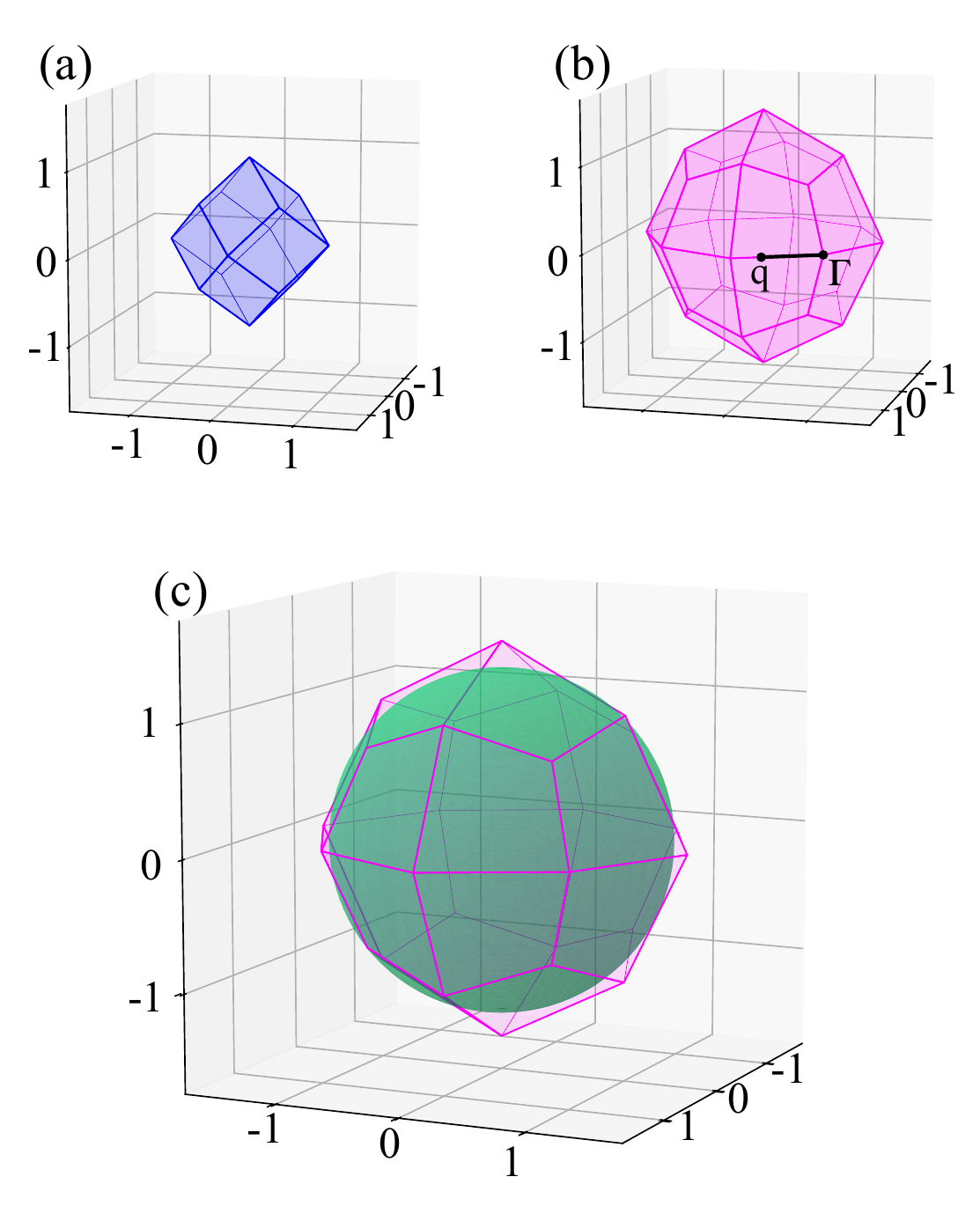}
   \caption{Zones in ${\bf k}$-space. (a) BCC Brillouin zone. (b) Jones' large zone formed by (2 1 1) planes and its equivalents. The $\Gamma$-${\rm q}$ path [see Fig.~\ref{fig:H3S-band-dos}] in the extended zone scheme is displayed. (c) Jones' zone overlaid by sphere with the same volume $V=9\times (2\pi/a)^3$.} 
   \label{fig:Zones}
   \end{center}
 \end{figure}

\section{Discussion: Jones zone mechanism}
We thus showed the plane-wave-like character of the electronic states in the compressed H$_{3}$S. The delocalization is enhanced by the nonlocal scattering of S ions due to the core electrons and the important DOS peak emerges by hybridization of multiple plane waves. The remaining question goes to why the DOS peak is near the Fermi level. As a solution to this, we propose a mechanism involving the large zone of Jones~\cite{Jones_book,Feng_JCP2010}. Jones have considered the effect of zones formed by Bragg planes distant from the origin in ${\bf k}$ space. The proximity of the free Fermi surface to one of such zone surfaces should promote hybridization of the plane waves at the zone surface by the corresponding Bragg diffraction, which lowers the electronic energy via band anticrossing. We notice that the present spherical distribution of $A({\bf k}+{\bf G};E_{\rm F})$ is significantly overlapping to the Jones' zone formed by $(2 1 1)$ and equivalent Bragg planes [Fig.~\ref{fig:Zones}(c)]. This $(2 1 1)$ zone [Fig.~\ref{fig:Zones}(b)] is close to the sphere and has a volume of $9\times (2\pi/a)^3$, which can exactly accommodate 9 valence electrons of H$_{3}$S per formula unit. On this $(2 1 1)$ zone surface, the diffraction due to the aligned S ions induces the hybridization of the plane waves. On its edges, multiple plane wave branches hybridize by the $(2 1 1)$ as well as $(1 1 0)$ and $(2 0 0)$ diffractions, which we remember correspond to $V_{3}, V_{1}$ and $V_{2}$, respectively. The hybridized band-anticrossing features are assured to be near the Fermi level due to the volume matching.

The hybridization at the Jones zone edges induces the saddle van Hove singularities in three dimensions~\cite{vanHove_PR1953,Akashi_PRB2020}. At the edge, the free band dispersion is increasing in the radial direction normal to the edge, whereas it is convex in the direction normal to both the edge and radius. Along the edge, it is also convex. The convex dispersions are relatively flat if the zone is large enough as they are tangents to the spherical equal energy surface. A weak hybridization at the edge yields a concave structure in the radial direction with the convex curves in the other directions intact, forming band saddle points linearly continued along the edge, which appear as the peaked concentration of the DOS. This general scenario applies to the H$_{3}$S case. Indeed, continuous saddle points run along the $\Gamma-q$ line~\cite{Akashi_Bragg_PRR2024}, whose flat dispersion induces the DOS peak (Fig.~\ref{fig:H3S-band-dos}(d)), and the $\Gamma-q$ line coincides with an edge of the $(2 1 1)$ zone as depicted in Fig.~\ref{fig:Zones}(b). 

Thus, the Jones zone mechanism explains why in this system the anomalous DOS peak emerges from the nearly uniform electrons and ensures that its location is near the Fermi level.

\section{Summary and conclusions}
The high-$T_{\rm c}$ superconductivity in the high-pressure H$_{3}$S has served an extreme case of the phonon-mediated superconductivity. For its $T_{\rm c}$, narrowly peaked DOS has been known of vital importance. The robust peak in the first-principles calculation has suggested a general mechanism to form such anomalous DOS structure. We concluded that all this originates from the interaction of plane waves. The first-principles KS wave functions were revealed to be significantly plane-wave-like. The complicated Fermi surface in the folded Brillouin zone expression is unfolded to be roughly spherical. Analysis on the local and nonlocal potentials yields effective plane-wave modeling with three or four parameters. The proximity to the Jones $(211)$ zone explains why the emerging DOS peak is located near the Fermi level, not because of fortune. 

In Ashcroft's hydride precompression scenario~\cite{Ashcroft_PRL2004}, transition of localized electrons to the common band states is an important precursor of metallic hydrogen-1$s$ states under relatively low pressure. There, atoms that accompany to H have been expected to provide extra valence electrons to the metallic band to increase the effective pressure. Our pseudopotential analysis suggests that S atoms serve another significant role on enhancing the delocalization via the core orthogonalization, which seems worth further studies for deriving insights into finding metallic hydrides at lower pressures.

Finally, delocalization of hydrogen-1$s$ electronic states is also important for suppressing the ferromagnetic spin-fluctuation and therefore boosting $T_{\rm c}$~\cite{Lu_KMgH3_PRB} in the hydrides. The Jones-zone scenario resolves the paradox of coexistence of narrow peak in the DOS, which would usually be related to localized orbitals with weak intersite hopping, and the delocalized plane-wave nature of the system. As the Jones zone mechanism in principle applies to general nearly uniform systems, this should help exploration of other high-pressure superconductors with the DOS-based boosting of $T_{\rm c}$.

\acknowledgments
This work was supported by JSPS KAKENHI Grant No. 23K03313 from the Japan Society for the Promotion of Science (JSPS). Part of the computation in this work has been done using the facilities of the Supercomputer Center, the Institute for Solid State Physics, the University of Tokyo (ISSPkyodo-SC-2024-Ba-0020, 2025-Ba-0047).

\bibliography{reference}

\end{document}